\documentclass{article}

\usepackage{arxiv}

\usepackage[utf8]{inputenc} 
\usepackage[T1]{fontenc}    
\usepackage{hyperref}       
\usepackage{url}            
\usepackage{booktabs}       
\usepackage{amsfonts}       
\usepackage{nicefrac}       
\usepackage{microtype}      
\usepackage{lipsum}
\usepackage{graphicx}
\graphicspath{ {./images/} }

\usepackage{algorithm}
\usepackage{algorithmic}

\usepackage{bibentry}

\usepackage{mathtools}

\usepackage{algorithm}
\usepackage{algorithmic}
\usepackage{amsfonts}
\usepackage{amsmath}
\usepackage{multirow}
\usepackage{enumitem}
\usepackage{newfloat}
\usepackage{listings}
\usepackage{graphicx}
\usepackage{hyperref}
\usepackage{balance}

\newtheorem{definition}{Definition}
\newtheorem{lemma}{Lemma}

\usepackage{array}

\usepackage{dsfont}
\usepackage[numbers]{natbib}

\date{}

\begin{document}

\title{Hybrid Advertising in the Sponsored Search}

\author{
    Zhen Zhang\thanks{Both authors contribute equally to this research.} \\
    \texttt{zhangzhen2023@ruc.edu.cn} \\
    Gaoling School of  \\ Artificial Intelligence, \\ Renmin University of China \\
    Beijing, China 
    \And
    Weian Li\footnotemark[1] \\
    \texttt{weian.li@sdu.edu.cn} \\
    School of Software, \\ Shandong University \\
    Jinan, China 
    \And
    \textbf{Yuhan Wang} \\
    \texttt{yuhanwang@ruc.edu.cn} \\
    Gaoling School of \\ Artificial Intelligence, \\ Renmin University of China \\
    Beijing, China 
    \And
    \\
    \textbf{Qi Qi}\thanks{Corresponding author.} \\
    \texttt{qi.qi@ruc.edu.cn} \\ 
    Gaoling School of \\ Artificial Intelligence, \\ Renmin University of China \\
    Beijing, China 
    \And
    \\
    \textbf{Kun Huang} \\
    \texttt{huangkun13@meituan.com}\\
    Meituan Inc. \\
    Beijing, China
}

\maketitle

\begin{abstract}
Online advertisements are a primary revenue source for e-commerce platforms. Traditional advertising models are store-centric, selecting winning stores through auction mechanisms. Recently, a new approach known as joint advertising has emerged, which presents sponsored bundles combining one store and one brand in ad slots. Unlike traditional models, joint advertising allows platforms to collect payments from both brands and stores. However, each of these two advertising models appeals to distinct user groups, leading to low click-through rates when users encounter an undesirable advertising model. To address this limitation and enhance generality, we propose a novel advertising model called ``Hybrid Advertising''. In this model, each ad slot can be allocated to either an independent store or a bundle. To find the optimal auction mechanisms in hybrid advertising, while ensuring nearly dominant strategy incentive compatibility and individual rationality, we introduce the Hybrid Regret Network (HRegNet), a neural network architecture designed for this purpose. Extensive experiments on both synthetic and real-world data demonstrate that the mechanisms generated by HRegNet significantly improve platform revenue compared to established baseline methods.
\end{abstract}

\keywords{Hybrid advertising; Sponsored search; Neural network; HRegNet}

\section{Introduction}
\label{sec:In}

Online advertising generates substantial revenue for internet companies and has become a primary revenue stream for giants like Google and Amazon. This revenue directly influences the growth of these companies, making the optimization of online ad revenue a key area of research. One common method for allocating online ad slots is the sponsored search auction. In these auctions, companies determine the allocation of ad slots and the prices advertisers should pay based on their bids and the established auction mechanism.

\begin{figure*}[ht]
    \centering
    \includegraphics[width=1\textwidth]{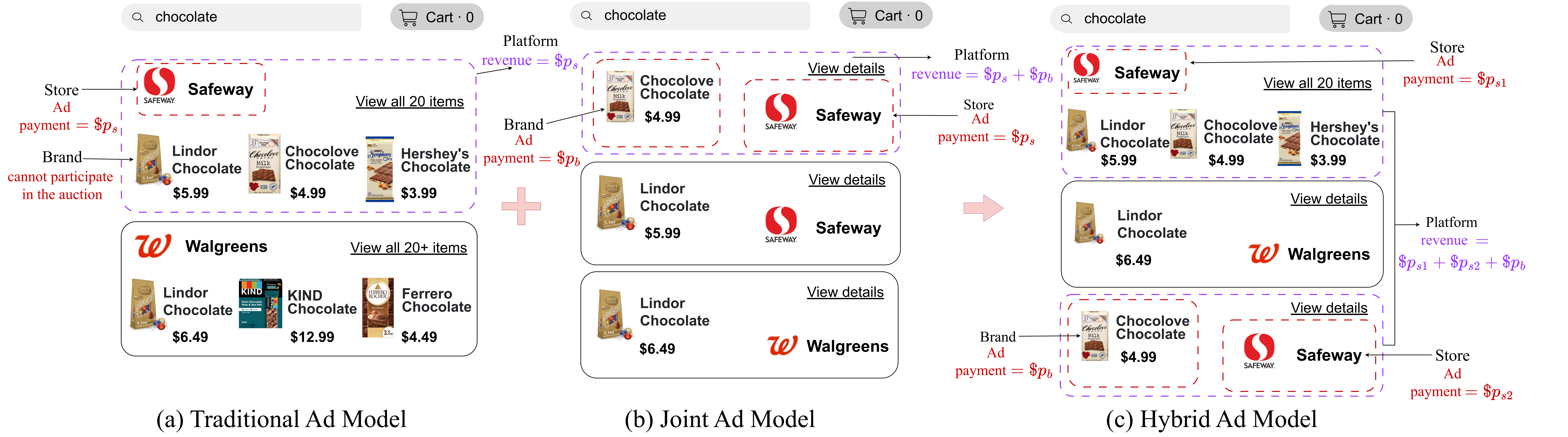}
    \caption{\label{fig:1}Traditional Ad Model, Joint Ad Model and Hybrid Ad Model.}
\end{figure*}

Taking e-commerce platforms like Instacart as an example, the traditional sponsored advertising model focuses on independent stores, where only stores can bid for ad slots to improve their visibility, as illustrated in Figure \ref{fig:1}(a).
Recently, a new online advertising model called joint advertising has emerged, allowing that both stores and brands participate in bidding for ad slots 
and each ad slot displays a bundle consisting of a store and a brand, as shown in Figure \ref{fig:1}(b). Compared to the traditional model, joint advertising generates revenue from both stores and brands, thus enhancing overall revenue, as supported by experiments \cite{zhang2024joint}. Many online e-commerce platforms, such as Meituan and Ele.me, have adopted this joint advertising model.

However, users might have different interest preferences for different advertisements \cite{zhou2018deep}, and the click-through rates (CTR) of different advertising models heavily depend on user preferences. For instance, when users are exploring potential products without a specific shopping target, a shop-centric display often generates higher CTRs than a bundle-centric display, as users tend to prefer broader options over specific recommendations. Conversely, when users know exactly what product they want to purchase, a bundle-centric display is more effective and attracts more clicks. This is because it not only provides an accurate list of results but also allows users to conveniently compare prices among options.
Statistical evidence from real datasets obtained from an online e-commerce platform indicates that the proportion of the former type of users is at least 60\%, 
while the latter type is less than 40\%. 
Given these findings, an important question arises: is it still a good choice for a platform to rely solely on joint advertising models? Moreover, is there an advertising model that can effectively balance traditional models with joint models?
To answer these questions, we propose a novel hybrid model that combines traditional advertising with joint advertising, as illustrated in Figure \ref{fig:1}(c). In this hybrid model, bidders can include both independent stores and brands, and a sponsored slot in the final displayed list can be allocated to either an independent store or a bundle comprising a store and a brand. In addition, the maximum number of displayed bundles can be flexibly controlled, to adapt the real-world requirement. We refer to this new model as ``hybrid advertising'' and find that it can generate more revenue than any existing advertising model.  

Meanwhile, hybrid advertising presents new challenges for auction mechanism design. First, since a store can appear either alone or within a bundle, designers must address the issue of correlated bids between independent stores and bundles, as well as among the bundles themselves. Second, the CTRs for independent stores and bundles may differ, necessitating careful consideration of which format to display in each slot. For example, there may exist a tradeoff that independent store bring more clicks while a bundle will gain more revenue. Lastly, when a bundle wins a slot, determining how to allocate payments between the store and brand while maintaining desirable auction properties poses a significant challenge.
These difficulties render current commonly used mechanisms unsuitable for hybrid scenarios. In recent years, despite theoretical challenges, rapid advancements in machine learning have led to the development of automated mechanism design (AMD) for computing optimal mechanisms \cite{dutting2019optimal,rahme2021permutation,duan2022context}. However, existing neural network architectures, even those that work well for joint advertising, are not well-suited for hybrid advertising, primarily because they cannot effectively address both allocation and pricing issues for independent stores and bundles simultaneously, given their different display effects. Therefore, the optimal design of hybrid mechanisms warrants in-depth research.



\subsection{Main Contributions}

In this paper, we introduce a novel and practical advertising sales model called the ``hybrid advertising'' model, where each ad slot can be allocated to either an independent store or a bundle of a store and a brand, and focus on a class of hybrid advertising model parameterized by the number of winning bundles, which can be flexibly applied to real world. We then propose a neural network architecture, \textbf{H}ybrid \textbf{Reg}ret \textbf{Net}work (HRegNet), designed to compute the optimal hybrid mechanisms that satisfy conditions of individual rationality (IR) and nearly dominant strategy incentive compatibility (DSIC). Finally, we conduct experiments on both synthetic and real datasets, demonstrating that the mechanisms generated by HRegNet can significantly increase revenue compared to established baselines.
To summarize, our main contributions are as follows:



\textbf{Hybrid auctions.} 
In the context of hybrid advertising, we refer to the auction mechanism as the ``hybrid auction''. In a hybrid auction, both stores and brands can participate in bidding. Unlike previous models, there are two types of winners: independent stores and bundles of stores and brands. Additionally, we limit the number of winning bundles to increase the generality and flexibility. 
The hybrid auction enhances the practicality of auction theory, potentially making a significant impact in both academia and industry.

\textbf{HRegNet.} 
To tackle the technical challenges mentioned earlier and achieve the optimal hybrid auction design, we propose an innovative neural network architecture called HRegNet. This architecture generates mechanisms that satisfy IR and near-DSIC, while maximizing revenue. HRegNet effectively addresses the unique challenges posed by hybrid advertising, which existing popular neural network architectures (e.g., \cite{feng2018deep, rahme2021permutation, duan2022context, zhang2024joint}) do not adequately resolve:

(1) \textbf{The correlated bids between independent stores and bundles, as well as within the bundles themselves.} 
Since each store and brand can belong to multiple bundles, and stores can also be displayed independently, correlated bids may arise among different bundles and stores. To address this issue, HRegNet inputs the bids from each store and brand and computes the allocation probability matrix for independent stores and bundles based on independent bids. This approach avoids relying on bundle bids, effectively mitigating the problem of correlated bids.

(2) \textbf{Two displayed forms of independent stores and bundles.}
A key feature of hybrid advertising is that both independent stores and bundles can be potential winners. Since different displayed forms may yield varying CTRs, it is essential to specify the winner along with its display form during slot allocation. To tackle this challenge, HRegNet utilizes two separate sub-neural networks within the allocation network to handle independent stores and bundles respectively, and then merges the allocation results to meet the necessary constraints, balancing the effects of stores and bundles. Note that even though JRegnet \cite{zhang2024joint} with necessary revision (details can be found in Section \ref{sec:exp}) can also be applied to hybrid setting, the structure of two separate sub-neural networks in HRegNet can capture the CTR difference of independent stores and bundles more effectively, leading a greater revenue shown in numerical experiments.



(3) \textbf{Implementation of hybrid auction constraints.} 
In hybrid auctions, several allocation constraints exist, such as limiting the number of winning bundles. Additionally, each independent store or bundle can occupy no more than one slot, and vice versa. To implement these constraints, HRegNet applies row-wise and column-wise softmax to the raw allocation matrix. It then incorporates the winning bundle constraint into a revised allocation matrix using a $C$-softmax function (assuming the number of winning bundles does not exceed $C$) to produce the final allocation.


(4) \textbf{Calculation of payments.} 
Determining payments for independent stores and bundles, as well as allocating payment between a store and brand within a bundle, is complex. HRegNet tackles this issue by training a series of parameters to scale each advertiser's total expected value, and define scaled values as their payments. Each parameter is calculated using a sigmoid function, ensuring that the payment is always less than the bid, and satisfying IR condition.

\textbf{Generality of Hybrid advertising and HRegNet.}
The hybrid advertising model can be extended to accommodate more general and complex scenarios. For example, slots can be allocated to independent brands or bundles; stores, brands, and bundles can be displayed exclusively within slots; or a bundle can consist of more than two components, provided the necessary bundle relationships exist. Furthermore, HRegNet can be adapted to these broader scenarios with only minor modifications.


\subsection{Related Work}
Classic auction mechanisms have been widely applied in online advertising systems. The Vickrey-Clarke-Groves (VCG) mechanism \cite{vickrey1961counterspeculation, clarke1971multipart, groves1973incentives} is designed to optimize the social welfare of advertisers while satisfying DSIC and IR. In contrast, the Myerson auction \cite{myerson1981optimal} focuses on optimizing revenue in single-parameter settings. Additionally, the generalized second price (GSP) mechanism \cite{edelman2007internet, caragiannis2011efficiency, gomes2009bayes, lucier2012revenue} and its variants \cite{lahaie2007revenue, thompson2013revenue} offer a computationally efficient and easily explainable model, leading to widespread adoption in current online platforms. However, due to the limitations of Myerson auctions caused by correlated bids and the challenges in defining GSP equilibrium, to the best of our knowledge, only the VCG mechanism is applicable in hybrid advertising contexts, but its revenue performance is not desirable. In this paper, we propose HRegNet to automatically compute an optimal mechanism that guarantees near-DSIC and IR.


With the advent of machine learning, the paradigm of automated mechanism design (AMD) \cite{conitzer2002complexity, sandholm2003automated, conitzer2004self, sandholm2015automated} has emerged to address the limitations of traditional mechanism design using algorithmic and data-driven methods. This approach leverages machine learning techniques to achieve approximately optimal auction designs \cite{balcan2008reducing, lahaie2011kernel, dutting2015payment}. 
The foundational neural network architecture for computing the optimal mechanisms that satisfy near-DSIC and IR in multi-item auctions is proposed by D{\"u}tting et al. \cite{dutting2019optimal, dutting2024optimal}. Subsequent research has expanded RegretNet to tackle various constraints and objectives in different auction scenarios, including fairness \cite{kuo2020proportionnet}, budget constraints \cite{feng2018deep}, human preferences \cite{peri2021preferencenet}, attention-based structure \cite{ivanov2022optimal}, anonymous symmetric auctions \cite{rahme2021permutation}, and contextual auctions \cite{duan2022context}. 
In the realm of joint auctions, Zhang et al. \cite{zhang2024joint} propose the JRegNet architecture for near-optimal joint auction design. Ma et al. \cite{ma2024joint} introduce the JAMA network based on the VCG mechanism to design joint auction mechanisms. Additionally, Aggarwal et al. \cite{aggarwal2024selling} examine joint advertising from a theoretical perspective, modeling it as a multi-period decision problem.
In this paper, we propose a novel and practical hybrid advertising model, and our HRegNet architecture is designed to address the challenges posed by hybrid auctions, enabling the optimal hybrid auction design.

\section{Model and Preliminaries}
\label{sec:HA}
In this section, we provide a detailed introduction to hybrid advertising and reformulate the optimal design of hybrid auctions as a learning problem.

\subsection{Hybrid Advertising Model}

In an e-commerce platform, consider that there are $m$ stores denoted by $M=\{1, 2, \cdots, m \}$ and $n$ brands denoted by $N=\{1, 2, \cdots, n \}$, participating in ad auctions for $K$ allocated ad slots. For each slot $k \in \{1, 2, \cdots, K\}$, let $\theta_k$ represent the position-related click-through rate (CTR), with the assumption that $1>\theta_1 \ge \cdots \ge \theta_K>0$. 

In the hybrid advertising model, each slot can display either an independent store or a bundle consisting of one store and one brand. A store can serve in two roles: as an independent entity or as part of a bundle with a brand\footnote{These two roles are not mutually exclusive. A store and its associated bundles can appear simultaneously in ad lists.}. Conversely, a brand must always be accompanied by a store, forming a bundle to be presented in a slot. Not all stores and brands can form bundles, and we use the indicator matrix $\{\mathbf{1}_{ij}\}_{i\in M, j\in N}$ to represent the relationships between stores and brands. If $\mathbf{1}_{ij} = 1$, it indicates a selling relationship between store $i$ and brand $j$, allowing them to be bundled during the auction stage; otherwise, no such relationship exists. It should be noted that for different auction samples under the same setting, $\{\mathbf{1}_{ij}\}_{i\in M, j\in N}$ is dynamically changing.
Additionally, to maintain the hybrid display, we restrict the number of bundles shown in an ad list to a maximum of $C$.

To differentiate the popularity of independent store displays and bundles, we assume that when store $i\in M$ is presented independently, a quality factor $\alpha_i > 0$ influences the CTR, reflecting the difference of independent display. Specifically, when store $i$ is displayed alone in slot $k$, the CTR is $\alpha_i \theta_k$. In contrast, if a bundle that includes store $i$ is allocated to slot $k$, the CTR is simply $\theta_k$. 

For each store $i \in M$ (or brand $j \in N$), we denote the value per click as $v_{i \cdot}$ (or $v_{\cdot j}$) which represents private information. We assume that the value $v_{i \cdot}$ (or $v_{\cdot j}$) is independently sampled from a distribution $F_{i \cdot}$ (or $F_{\cdot j}$) within the domain $V_{i \cdot}$ (or $V_{\cdot j}$). The value profile is denoted by $\boldsymbol{v} = (v_{1 \cdot}, \cdots, v_{m \cdot}, v_{\cdot 1}, \cdots, v_{\cdot n}) \in \mathbb{V} = V_{1 \cdot} \times \cdots \times V_{m \cdot} \times V_{\cdot 1} \times \cdots \times V_{\cdot n}$. Additionally, let $\boldsymbol{v}_{-i \cdot} = (v_{1 \cdot}, \cdots, v_{(i-1) \cdot}, v_{(i+1) \cdot}, \cdots, v_{\cdot n})$ be the value profile excluding store $i$, with a similar notation used for brand $j$. 

We now formally establish a hybrid auction mechanism. Prior to the auction, each store $i$ and each brand $j$ independently submit bids $b_i$ and $b_j$ for each click, which may differ from their respective values $v_i$ and $v_j$. Similarly, we denote $\boldsymbol{b}$, $\boldsymbol{b}_{-i \cdot}$ and $\boldsymbol{b}_{\cdot -j}$ by the relevant bid profiles, respectively. A hybrid auction mechanism $\mathcal{M}(g,p)$ consists of an allocation rule $g=\big((g_{i \cdot})_{i\in M}, (g_{\cdot j})_{j\in N} \big)$ and a payment rule $p=\big((p_{i \cdot})_{i\in M}, (p_{\cdot j})_{j\in N} \big)$. In the allocation rule, $g_{i \cdot}(\boldsymbol{b}) = \sum_{k=1}^{K}{a_{i \cdot k}(\boldsymbol{b}) \theta_k}$ represents the expected CTR that store $i$ can achieve, where $a_{i \cdot k}(\boldsymbol{b})$ is the total probability that store $i$ can be allocated to slot $k$, regardless of its display form\footnote{Herein, we incorporate $\alpha_i$ into $a_{i \cdot k}(\boldsymbol{b})$, when calculating the probability for store $i$ displayed independently.}. A similar definition applies to brand $j$. The payment rule $p_{i \cdot}$ (or $p_{\cdot j}$) $: \mathbb{V} \rightarrow \mathbb{R}_{\geq 0}$ represents the payment that store $i$ (or brand $j$) is required to charge. 

In this model, we focus on general requirements for the ad auction: one slot can be allocated to either one independent store or one bundle, and one bundle or one independent store can occupy at most one slot. Additionally, in the hybrid auction, the number of slots allocated to bundles must not exceed $C$. We refer to these constraints as feasibility conditions.

Given the definition of a hybrid auction, the utility of each store $i$ can be expressed in a quasi-linear form: $u_{i \cdot}(v_{i \cdot} ; \boldsymbol{b}) = v_{i \cdot}(g_{i \cdot}(\boldsymbol{b})) - p_{i \cdot}(\boldsymbol{b})$. A similar utility definition applies to each brand $j$. Each advertiser, including both stores and brands, aims to maximize their own utility.

This paper examines two desirable properties of mechanisms: dominant strategy incentive compatibility (DSIC) and individual rationality (IR). DSIC ensures that participants have no incentive to misreport their bids, meaning that bidding truthfully yields the highest utility. IR guarantees that each participant's utility is non-negative. Definitions of DSIC and IR can be found in Definition \ref{def:DSIC} and Definition \ref{def:IR}, respectively.

\begin{definition}[DSIC]\label{def:DSIC}
A hybrid auction satisfies dominant strategy incentive compatibility if any store $i$ (or brand $j$) maximizes its utility by reporting its true valuation, regardless of the reports of other participants. This can be expressed as follows:
$$u_{i \cdot}(v_{i \cdot}; (v_{i \cdot}, \boldsymbol{b}_{-i \cdot})) \ge u_{i \cdot}(v_{i \cdot}; (b_{i \cdot}, \boldsymbol{b}_{-i \cdot})),$$ 
$$ \forall i \in M, \forall v_{i \cdot} \in V_{i \cdot}, \forall b_{i \cdot} \in V_{i \cdot}, \forall \boldsymbol{b}_{-i \cdot} \in \mathbb{V}_{-i \cdot},$$ and 
$$u_{\cdot j}(v_{\cdot j}; (v_{\cdot j}, \boldsymbol{b}_{-\cdot j})) \\ \ge u_{\cdot j}(v_{\cdot j}; (b_{\cdot j}, \boldsymbol{b}_{-\cdot j})),$$ $$\forall j \in N, \forall 
 v_{\cdot j} \in V_{\cdot j}, \forall  b_{\cdot j} \in V_{\cdot j}, \forall  \boldsymbol{b}_{-\cdot j} \in \mathbb{V}_{-\cdot j}.$$
\end{definition}

\begin{definition}[IR]\label{def:IR}
A hybrid auction satisfies individual rationality if any store $i$ (or brand $j$) receives a non-negative utility by reporting truthfully. In other words, this can be expressed as:
$$u_{i \cdot}(v_{i \cdot}; (v_{i \cdot}, \boldsymbol{b}_{-i \cdot})) \ge 0, \quad \forall i \in M, \forall v_{i \cdot} \in V_{i \cdot}, \forall \boldsymbol{b}_{-i \cdot} \in \mathbb{V}_{-i \cdot},$$ and $$u_{\cdot j}(v_{\cdot j}; (v_{\cdot j}, \boldsymbol{b}_{-\cdot j})) \ge 0, \quad \forall j \in N, \forall v_{\cdot j} \in V_{\cdot j}, \forall \boldsymbol{b}_{-\cdot j} \in \mathbb{V}_{-\cdot j}.$$
\end{definition}

In a DSIC and IR hybrid auction, the expected revenue is defined as:
$$ rev := \mathbb{E}_{\boldsymbol{v} \sim F}\big[\sum_{i=1}^{m}{p_{i \cdot}(\boldsymbol{v})} + \sum_{j=1}^{n}{p_{\cdot j} (\boldsymbol{v})} \big],$$
where $F$ represents the joint value distribution of all stores and brands. The objective of the optimal hybrid auction design is to find a feasible auction mechanism that maximizes the expected revenue while ensuring DSIC and IR.

\subsection{Hybrid Auction Design as a Learning Problem}

The problem of the optimal hybrid auction design can be formulated as a learning problem. To achieve this, we first introduce the concept of ex-post regret. The ex-post regret for store $i$ is defined as: 
$$
rgt_{i \cdot}(\boldsymbol{v})=\mathbb{E}_{v \sim F}[\max_{v_{i \cdot}' \in V_{i \cdot}}{u_{i \cdot} (v_{i \cdot}; (v_{i \cdot}', \boldsymbol{v}_{-i \cdot}))} - u_{i \cdot} (v_{i \cdot};\boldsymbol{v})] \text{\hfill .}
$$
Specifically, with the bids of others fixed, the ex-post regret for store $i$ represents the maximum increase in utility that it can achieve by misreporting its bid. A similar definition applies to brand $j$. 

Then, an auction mechanism satisfies DSIC if and only if, for all stores and brands, $rgt_{i \cdot}(\boldsymbol{v}) = 0$ and $rgt_{\cdot j}(\boldsymbol{v}) = 0$. Furthermore, we formulate the problem of the optimal hybrid auction design as a constrained optimization problem:
\begin{align}
\label{jj2}
\min_{(g,p) \in \mathcal{M}} \quad & -\mathbb{E}_{v \sim F}[\sum_{i=1}^{m}{p_{i \cdot}(\boldsymbol{v})} + \sum_{j=1}^{n}{p_j(\boldsymbol{v})}] \\
\text {s.t.} \quad & rgt_{i \cdot}(\boldsymbol{v}) = 0, \quad i = 1, \cdots, m,  \notag \\
& rgt_{\cdot j}(\boldsymbol{v}) = 0, \quad j = 1, \cdots, n \text{\hfill .} \notag
\end{align}
where $\mathcal{M}$ represents the set of all feasible hybrid auction mechanisms that ensure all bundles obtain at most $C$ slots in total and satisfy the IR condition. Due to the intricate constraints, this optimization problem is often intractable \cite{conitzer2002complexity, conitzer2004self}. To address this issue, we parameterize the auction mechanisms using $w \in \mathbb{R}^{x}$ as $\mathcal{M}^w (g^w, p^w) \subseteq \mathcal{M}(g, p)$. This allows us to optimize the parameter 
$w$ to find the optimal mechanism $\mathcal{\tilde{M}}^w (g^w, p^w) \in \mathcal{M}^w (g^w, p^w)$ that maximizes expected revenue while satisfying DSIC and IR.

Using $L$ samples of value profiles independently drawn from the joint distribution $F$, we compute empirical terms to estimate all expected values. For $\mathcal{M}^w (g^w, p^w)$, the empirical ex-post regret for each store $i$ (and similar for brand $j$) is calculated as follows:
\begin{align}
\label{j2}
\widehat{r g t}_{i \cdot}(w)  =   \frac{1}{L} \sum_{l=1}^{L}[\max_{v_{i \cdot}^{\prime} \in V_{i \cdot}} u_{i \cdot}^w (v_{i \cdot}^{(l)}; (v_{i \cdot}^{\prime}, \boldsymbol{v}_{-i \cdot}^{(l)}))  
  - u_{i \cdot}^w (v_{i \cdot}^{(l)}; \boldsymbol{v}^{(l)})] \text{\hfill .}
\end{align}
Based on this, the optimization problem (\ref{jj2}) can be expressed as follows:
\begin{align}
\label{j1}
\min_{w \in \mathbb{R}^{d}} \quad & -\frac{1}{L} \sum_{l=1}^{L}{[\sum_{i=1}^{m}{p_{i \cdot}^w(\boldsymbol{v}^{(l)})} + \sum_{j=1}^{n}{p_{\cdot j}^w(\boldsymbol{v}^{(l)})}]} \\
\text {s.t.} \quad & \widehat{rgt}_{i \cdot}(w) = 0, \quad i = 1, \cdots, m, \notag\\
& \widehat{rgt}_{\cdot j}(w) = 0, \quad j = 1, \cdots, n \text{\hfill .} \notag
\end{align}
The constraints that all bundles can obtain at most $C$ slots in total, along with the IR condition, are enforced through the network architecture detailed in Section \ref{sec:HR}.

\section{HRegNet}
\label{sec:HR}
In this section, after transforming the optimal hybrid auction design problem into a learning problem, we propose \textbf{H}ybrid \textbf{Reg}ret \textbf{Net}work (HRegNet) to generate mechanisms that maximize the expected revenue while ensuring near-DSIC and IR.

\begin{figure*}[ht]
    \centering
    \includegraphics[width=1\textwidth]{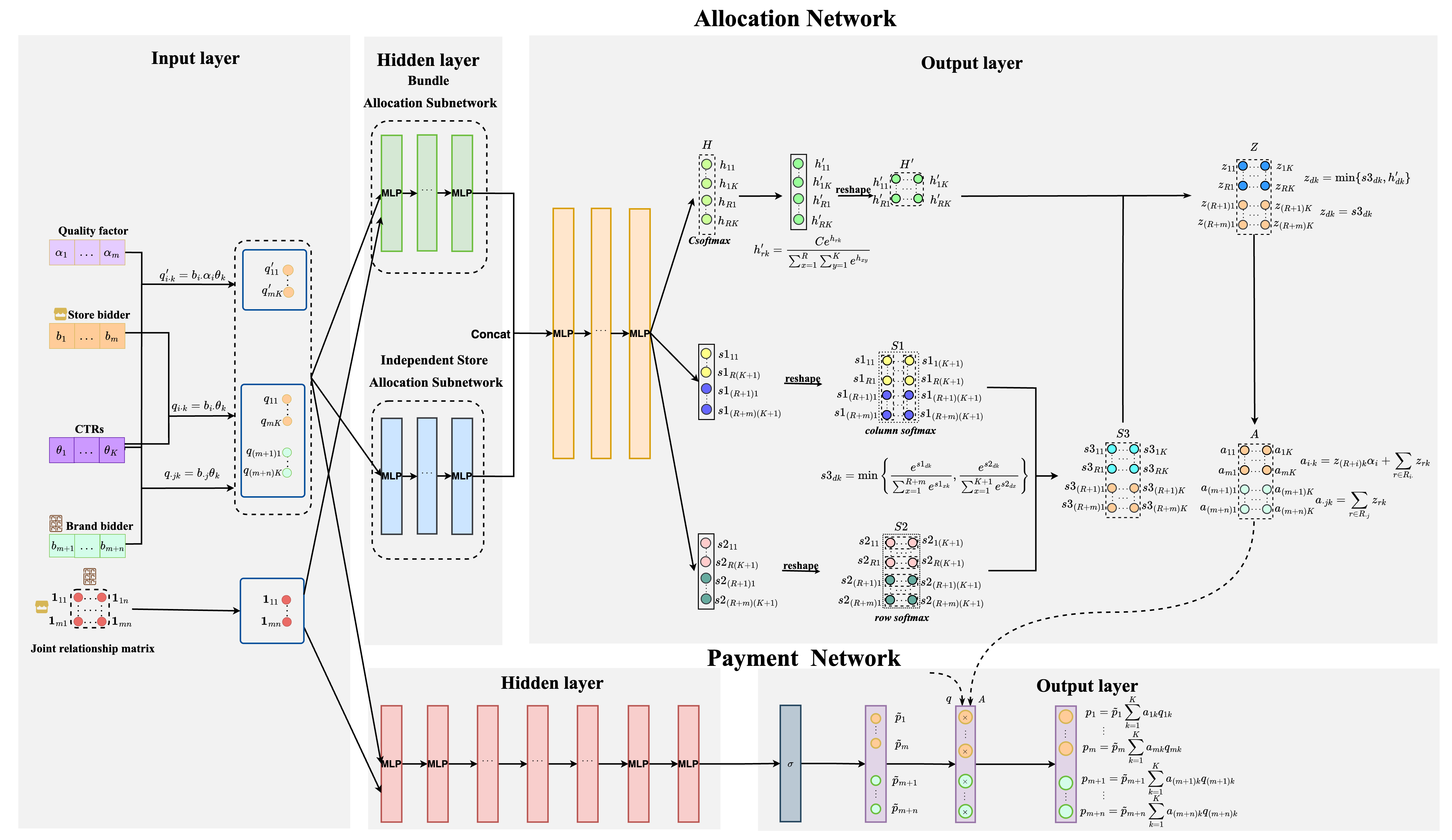}
    \caption{\label{fig:3} The architecture of HRegNet is designed for a scenario with $m$ stores, $n$ brands, $K$ slots, and $R$ bundles. The maximum number of winning bundles is $C$.} 
\end{figure*}

\subsection{The HRegNet Architecture}
The architecture of HRegNet is shown in Figure \ref{fig:3}. There are two components in HRegNet: the allocation network and the payment network. The allocation network also consists of two parts: the independent store allocation subnetwork and the bundle allocation subnetwork. Next, we elaborate on the components of the HRegNet architecture.

\textbf{Input of HRegNet.} The inputs of the bundle allocation subnetwork and the payment network comprise the joint relationship matrix $\{\mathbf{1}_{ij}\}_{i\in M, j\in N}$, and the expected bids from store $i$ (as part of a bundle) and brand $j$ for slot $k$, $q_{i \cdot k}$ and $q_{\cdot j k}$, expressed respectively as:
\begin{equation*}
\left\{\begin{array}{ll}
q_{i \cdot k}=\theta_{k} b_{i \cdot}, & \forall i \in M, ~ \forall k \in \{1,\cdots,K \} \text{\hfill ,}\\
q_{\cdot j k}=\theta_{k} b_{\cdot j}, & \forall j \in N, ~ \forall k \in \{1,\cdots,K \} \text{\hfill ,}
\end{array}\right.
\label{me1}
\end{equation*}
The input of the independent store allocation subnetwork is the expected bid  $q'_{i \cdot k}$ from independent store $i$ for slot $k$, calculated by:
$$
q'_{i \cdot k}=\theta_{k} b_{i \cdot} \alpha_i,\quad \forall i \in M, ~ \forall k \in \{1,\cdots,K \}. 
$$

\textbf{Calculating the allocation probability matrix $Z$ of independent stores and bundles.} 
Recall that the number of stores is $m$ and denote $R$ by the number of bundles involved in the hybrid auction. Now, we elaborate how to compute the matrix $Z$. Matrix $Z$ consists of two parts: the allocation probabilities of all bundles denoted by $\{z_{rk}\}_{r\in \{1,2,\cdots, R\}, k\in \{1,\cdots,K\}}$, and the allocation probabilities of independent stores denoted by $\{z_{ik}\}_{i\in M, k\in \{1,\cdots,K\}}$, as shown by the blue and orange matrix in Figure \ref{fig:3}, respectively.
 
To calculate matrix $Z$, as shown in Figure \ref{fig:3}, we take the results of hidden layer as the input, 
and mainly concentrate on realizing the feasible constraints introduced in Section \ref{sec:HA}: (a) a slot can be allocated to at most one independent store or one bundle, i.e., $\sum_{r=1}^{R}  {z}_{r k} + \sum_{i=1}^{m}  {z}_{i k} \leq 1, \forall k \in \{1,\cdots,K\}$; (b) a bundle or an independent store can obtain at most one slot, i.e., $\sum_{k=1}^{K}  {z}_{r k} \leq 1, \forall r \in \{1,\cdots,R\}$ and  $\sum_{k=1}^{K}  {z}_{i k} \leq 1,  \forall i \in M$; (c) all bundles can obtain at most $C$ slots, i.e., $ \sum_{r=1}^{R} \sum_{k=1}^{K} {z}_{r k} \leq C$. We incorporate all above constraints into the construction process of the allocation probability matrix $Z$ and introduce allocation network from left to right. First, we reshape the results of hidden layer to get matrices $S1$ and $S2$ to adapt the dimension of slots. 
Herein, note that matrices $S1$ and $S2$ have the same scale, where the row represents an independent store or a bundle and the column represents a slot or non-allocation. Next, 
we execute column-wise (and row-wise) normalization on $S1$ (and $S2$) by a softmax function, and define $S3$ by taking the minimum element at each corresponding position of normalized $S1$ and $S2$, ignoring the column of non-allocation, i.e., 
\begin{align*}
s3_{d k}=\varphi_{d k}^{B S}\left(S1, S2\right)=\min \left\{\frac{e^{s1_{d k}}}{\sum_{x=1}^{R+m} e^{s1_{x k}}}, \frac{e^{s2_{d k}}}{\sum_{x=1}^{K+1} e^{s2_{d x}}}\right\} \text{\hfill ,}
\end{align*}
for all $d \in \{1, \cdots,R,R+1,\cdots,R+m\}$ and $k \in \{1, \cdots,K\}$, 
where $s1_{dk}$ and $s2_{dk}$ are the elements of $S1$ and $S2$, respectively. We show that matrix $S3$ is a doubly stochastic matrix by the Lemma \ref{lemma1}, which means that constraints (a) and (b) are satisfied. 
\begin{lemma}[\cite{dutting2019optimal}]
\label{lemma1}
The matrix $\varphi^{B S}\left(S1, S2\right)$ is doubly stochastic for all $S1, S2 \in \mathbb{R}^{(R+m) K}$. For any doubly stochastic matrix $S3 \in[0,1]^{(R+m) K}$, there exist $S1, S2 \in \mathbb{R}^{(R+m) K}$ such that $S3=\varphi^{B S}\left(S1, S2\right)$.
\end{lemma}

After obtaining matrix $S3$, to further satisfy constraint (c), we apply softmax normalization to matrix $H$ and multiply each element by $C$ to obtain matrix $H'$.
In detail, for any $r\in \{1,\cdots,R \}$ and $ k \in \{1,\cdots,K \}$, $h'_{rk}$ can be calculated by 
    \begin{align*}
        h'_{rk} = \frac{C e^{h_{r k}}}{\sum_{x=1}^{R} \sum_{y=1}^{K} e^{h_{x y}}},
    \end{align*}
where $h_{rk}$ is the element of matrix $H$. Since the sum of all elements of matrix $H'$ is $C$, matrix $H'$ satisfies constraint (c).
Finally, we combine matrices $S3$ and $H'$ to get matrix $Z$, by the following rule, 
\begin{align*}
\left\{\begin{array}{l}
z_{d k}= \min\{s3_{d k}, h'_{d k}\}, \quad \forall d \in \{1,\cdots,R \}, ~ \forall k \in \{1,\cdots,K \} \text{\hfill ,}\\
z_{d k}= s3_{d k}, \quad \quad \forall d \in \{R+1,\cdots,R+m \}, ~ \forall k \in \{1,\cdots,K \} \text{\hfill .}
\end{array}\right.
\end{align*}
Intuitively, we add the constraint (c) in matrix $S3$ by comparing the corresponding value in matrix $H'$. Therefore, we get the allocation probability matrix $Z$, which satisfies constraints (a), (b), and (c), simultaneously. See the allocation network in Figure \ref{fig:3} for the entire process.

\textbf{Transforming the allocation probability matrix $Z$ into the allocation probability matrix $A$.} 
Probability matrix $A$ represents the allocation probability of store $i$ (or brand $j$) to slot $k$. Mathematically, the probability of allocating store $i$ to slot $k$ is equal to the sum of the allocation probability of independent store $i$ in slot $k$ multiplied by its quality factor $\alpha_i$ and the allocation probabilities of all bundles including it in slot $k$. The allocation probability for brand $j$ in slot $k$ is similar, but does not include the case of independent brand. Therefore, matrix $A$ can be expressed as 
\begin{align*}
\left\{\begin{array}{ll}
a_{i \cdot k}= z_{(R+i) k} \alpha_i + \sum_{r \in R_{i \cdot}} z_{r k}, & \forall i \in M,  ~  \forall k \in \{1,\cdots,K \} \text{\hfill ,}\\
a_{\cdot j k}=\sum_{r \in R_{\cdot j}} z_{r k}, &\forall j \in N, ~ \forall k \in \{1,\cdots,K \} \text{\hfill ,}
\end{array}\right.
\end{align*}
where $R_{i \cdot}$ and $R_{\cdot j}$ represent the sets of all bundles including store $i$ and brand $j$, respectively. 

\textbf{Calculating the payment matrix $P$.} Having the allocation matrix $A$, we can propagate it to the payment network to compute the payment matrix $P$. In the payment matrix $P$, the first $m$ rows correspond to the payments $p_i$ for each store $i$, and the last $n$ rows correspond to the payments $p_j$ for each brand $j$, as shown by the orange and pale green matrix in Figure \ref{fig:3}, respectively. The payments $p_i$ and $p_j$ are presented as follows:
\begin{align*}
\left\{\begin{array}{ll}
p_{i \cdot}=\tilde{p}_{i \cdot}\left(\sum_{k=1}^{K} a_{i \cdot k} q_{i \cdot k}\right), & \forall i \in M \text{\hfill ,} \\
p_{\cdot j}=\tilde{p}_{\cdot j}\left(\sum_{k=1}^{K} a_{\cdot j k} q_{\cdot j k}\right), & \forall j \in N \text{\hfill ,}
\end{array}\right.
\end{align*}
where $\tilde{p}_{i \cdot}, \tilde{p}_{\cdot j} \in [0,1]$ are parameters calculated by the sigmoid function, to scale the expected values of store $i$ and brand $j$ obtained in hybrid auctions. Since we focus on DISC mechanisms, and $\tilde{p}_{i \cdot} \in [0,1]$ and $\tilde{p}_{\cdot j} \in [0,1]$, the payments cannot exceed the values (i.e., bids), which implies that IR condition is satisfied.


\subsection{Training of HRegNet} 
In this subsection, we provide a detailed introduction on how to train HRegNet.

\textbf{Transformation of the optimization objective for obtaining the objective function.} The augmented Lagrangian method is used to convert the constrained optimization problem (\ref{j1}) into an unconstrained optimization, which is the objective function of HRegNet, as follows:
\begin{align*}
\mathcal{C}_{\rho}(w ; \boldsymbol{\lambda})= &-\frac{1}{L} \sum_{\ell=1}^{L}\left[\sum_{i=1}^{m} p_{i \cdot}^{w}\left(\boldsymbol{v}^{(\ell)}\right)+\sum_{j=1}^{n} p_{\cdot j}^{w}\left(\boldsymbol{v}^{(\ell)}\right)\right]  \\
& + \sum_{i=1}^{m}{\lambda_{i \cdot} \widehat{rgt}_{i \cdot}(w)} + \sum_{j=1}^{n}{\lambda_{\cdot j} \widehat{rgt}_{\cdot j}(w)} \\
& + \frac{\rho}{2}\sum_{i=1}^{m}{(\widehat{rgt}_{i \cdot}(w))^2} +\frac{\rho}{2} \sum_{j=1}^{n}{(\widehat{rgt}_{\cdot j}(w))^{2}} \text{\hfill ,}
\end{align*}
where $\boldsymbol{\lambda} \in \mathrm{R}^{m+n}$ denotes the Lagrange multiplier and $\rho>0$ represents the penalty factor.

Afterwards, we divide the training set into minibatches with $E$ auction samples, and we take a minibatch of samples $\mathcal{E}_t = \{v^{(1)}, \cdots, v^{(E)}\}$ for training in each iteration $t \in \{1, \cdots ,T\}$, where $T$ is the total number of iterations.

\textbf{Computing the optimal misreports.} To train HRegNet, in addition to the objective function and the division of the training set, it is also necessary to find the optimal misreports to calculate the regret. The gradient ascent method is used to find the optimal misreports: 
\begin{align*}
\left\{\begin{array}{l}
v_{i \cdot}^{\prime(\ell)}=v_{i \cdot}^{\prime(\ell)}+\left.\tau \nabla_{v_{i \cdot}^{\prime}}\left[u_{i \cdot}^{w}\left(v_{i \cdot}^{(\ell)} ;\left(v_{i \cdot}^{\prime}, \boldsymbol{v}_{-i \cdot}^{(\ell)}\right)\right)\right]\right|_{v_{i \cdot}^{\prime}=v_{i \cdot}^{\prime(\ell)}} \text{\hfill ,}\\
v_{\cdot j}^{\prime(\ell)}=v_{\cdot j}^{\prime(\ell)}+\left.\tau \nabla_{v_{\cdot j}^{\prime}}\left[u_{\cdot j}^{w}\left(v_{\cdot j}^{(\ell)} ;\left(v_{\cdot j}^{\prime}, \boldsymbol{v}_{-\cdot j}^{(\ell)}\right)\right)\right]\right|_{v_{\cdot j}^{\prime}=v_{\cdot j}^{\prime(\ell)}} \text{\hfill ,}
\end{array}\right.
\end{align*}
for some $\tau>0$.

After obtaining the regret, the objective function ${C}_{\rho}(w ; \boldsymbol{\lambda})$ can be computed, thereby performing backpropagation to update the network parameters $w$, and thus minimize the objective function ${C}_{\rho}(w ; \boldsymbol{\lambda})$.

\textbf{Updating the Lagrange multipliers.} During the training process, at fixed intervals of iterations, the Lagrange multipliers are updated through:
\begin{align*}
\left\{\begin{array}{l}
\lambda_{i \cdot }^{t+1}=\lambda_{i \cdot }^{ {t }}+\rho_{t} \widetilde{r g t}_{i \cdot }\left(w^{ {t+1 }}\right) , \quad \forall i \in M\text{\hfill ,}\\
\lambda_{\cdot j }^{ {t+1 }}=\lambda_{\cdot j }^{ {t }}+\rho_{t} \widetilde{r g t}_{\cdot j}\left(w^{ {t+1 }}\right) , \quad \forall j \in N \text{\hfill ,}
\end{array}\right.
\end{align*}
where $\widetilde{r g t}_{i \cdot}(w)$ and $\widetilde{r g t}_{\cdot j}(w)$ denote the empirical regret for minibatch $\mathcal{E}_t$ computed based on Equation (\ref{j2}). The complete algorithm of training HRegNet is shown in Algorithm $1$.

\begin{algorithm}[h!]
\caption{HRegNet Training} 
\begin{algorithmic}[1]
\STATE {\bf Input:} Minibatches 
$\mathcal{E}_1, . . . , \mathcal{E}_T$ of size E \\
\STATE {\bf Parameters:} 
$\forall t \in \{1,\ldots,T\}, \rho_t > 0, \alpha > 0, \eta > 0, \Gamma \in \mathbb{N}, T \in \mathbb{N}, \gamma \in \mathbb{N}
$ \\
\STATE {\bf Initialize:} 
$w^0 \in \mathbb{R}^{x}, \lambda^0 \in \mathbb{R}^{m+n}$
\FOR{$t=0$ to $T$} 
  \STATE Receive minibatch $\mathcal{L}_t = \{v^{(1)}, \ldots, v^{(E)}\}$ 
  \STATE Initialize misreport $\forall \ell \in \{1,\ldots,E\}, v_{i \cdot}^{\prime(\ell)} \in V_{i \cdot},v_{\cdot j}^{\prime(\ell)} \in V_{\cdot j}, i \in M,  j \in N$
  \FOR{$r=0$ to $\Gamma$} 
    \STATE $\forall \ell \in \{1,\ldots,E\}, i \in M,j \in N:$
     \STATE $\quad$ $
            v_{i \cdot}^{\prime(\ell)}=v_{i \cdot}^{\prime(\ell)}+\left.\tau \nabla_{v_{i \cdot}^{\prime}}\left[u_{i \cdot}^{w}\left(v_{i \cdot}^{(\ell)} ;\left(v_{i \cdot}^{\prime}, \boldsymbol{v}_{-i \cdot}^{(\ell)}\right)\right)\right]\right|_{v_{i \cdot}^{\prime}=v_{i \cdot}^{\prime(\ell)}} $
      \STATE $\quad$ $
          v_{\cdot j}^{\prime(\ell)}=v_{\cdot j}^{\prime(\ell)}+\left.\tau \nabla_{v_{\cdot j}^{\prime}}\left[u_{\cdot j}^{w}\left(v_{\cdot j}^{(\ell)} ;\left(v_{\cdot j}^{\prime}, \boldsymbol{v}_{-\cdot j}^{(\ell)}\right)\right)\right]\right|_{v_{\cdot j}^{\prime}=v_{\cdot j}^{\prime(\ell)}}  $
  \ENDFOR
  \STATE Compute regret gradient:$\forall \ell \in \{1,\ldots,E\}, i \in M, j \in N:$
  \STATE $\quad$ $g_{\ell, i \cdot}^{t}= \left. \nabla_{w}\left[u_{i \cdot}^{w}\left(v_{i \cdot}^{(\ell)} ;\left({v^{\prime}}_{i \cdot}^{(\ell)}, \boldsymbol{v}_{-i \cdot}^{(\ell)}\right)\right)-u_{i \cdot}^{w}\left(v_{i \cdot}^{(\ell)} ; \boldsymbol{v}^{(\ell)}\right)\right]\right|_{w=w^{t}}$
  \STATE $\quad$ $g_{\ell, \cdot j}^{t}= \left. \nabla_{w}\left[u_{\cdot j}^{w}\left(v_{\cdot j}^{(\ell)} ;\left({v^{\prime}}_{\cdot j}^{(\ell)}, \boldsymbol{v}_{-\cdot j}^{(\ell)}\right)\right)-u_{\cdot j}^{w}\left(v_{\cdot j}^{(\ell)} ; \boldsymbol{v}^{(\ell)}\right)\right]\right|_{w=w^{t}}$
  \STATE Use Formula (\ref{eueu}) to calculate Lagrangian gradient and update $w^t$:
  \STATE $\quad$ $w^{t+1} \leftarrow w^{t}- \eta \nabla_{w} \mathcal{C}_{\rho_{t}}\left(w^{t}, \lambda^{t}\right)$
  \STATE Update Lagrange multipliers every $\gamma$ iterations:
    \STATE $\quad$ {\bf if} {$t$ is an integer multiple of  $\gamma$} {\bf then} 
    \STATE $\quad$  $\quad$  $\lambda_{i \cdot}^{t+1} \leftarrow \lambda_{i \cdot}^{t}+\rho_{t} \widetilde{r g t}_{i \cdot}\left(w^{t+1}\right), \quad \forall i \in M$
     \STATE $\quad$  $\quad$  $\lambda_{\cdot j}^{t+1} \leftarrow \lambda_{\cdot j}^{t}+\rho_{t} \widetilde{r g t}_{\cdot j}\left(w^{t+1}\right), \quad \forall j \in N$
  \STATE $\quad$ {\bf else}
    \STATE $\quad$  $\quad$ $\boldsymbol{\lambda}^{t+1} \leftarrow \boldsymbol{\lambda}^t$
   \STATE $\quad$ {\bf end if}
\ENDFOR
\end{algorithmic}
\end{algorithm}

For fixed $\boldsymbol{\lambda}^t$, w.r.t. $w$, the gradient of $C_\rho$ is formulated as:

\begin{align}
\label{eueu}
\quad \nabla_{w} \mathcal{C}_{\rho}\left(w ; \boldsymbol{\lambda}^{t}\right)= &-\frac{1}{E} \sum_{\ell=1}^{E} \left[\sum_{i=1}^{m} \nabla_{w} p_{i \cdot }^{w}\left(\boldsymbol{v}^{(\ell)}\right)+ \sum_{j=1}^{n} \nabla_{w} p_{\cdot j}^{w}\left(\boldsymbol{v}^{(\ell)}\right)\right] \notag\\
& + \sum_{\ell=1}^{E}\left[ \sum_{i=1}^{m}  \lambda_{i \cdot}^{t} g_{\ell, i \cdot}+\sum_{j=1}^{n} \lambda_{\cdot j}^{t} g_{\ell, \cdot j} \right] + \notag\\
& \rho \sum_{\ell=1}^{E} \left[\sum_{i=1}^{m}  \widetilde{r g t}_{i \cdot}(w) g_{\ell, i \cdot }+ \sum_{j=1}^{n} \widetilde{r g t}_{\cdot j}(w) g_{\ell, \cdot j}\right] 
\end{align}

where

\begin{align*}
\quad \quad
\left\{\begin{array}{l}
g_{\ell, i \cdot}=\nabla_{w}\left[\max _{v_{i \cdot}^{\prime} \in V_{i \cdot}} u_{i \cdot}^{w}\left(v_{i \cdot}^{(\ell)} ;\left(v_{i \cdot}^{\prime}, \boldsymbol{v}_{-i \cdot}^{(\ell)}\right)\right)-u_{i \cdot}^{w}\left(v_{i \cdot}^{(\ell)} ; \boldsymbol{v}^{(\ell)}\right)\right]  \text{\hfill ,}\\
g_{\ell, \cdot j}=\nabla_{w}\left[\max _{v_{\cdot j}^{\prime} \in V_{\cdot j}} u_{\cdot j}^{w}\left(v_{\cdot j}^{(\ell)} ;\left(v_{\cdot j}^{\prime}, \boldsymbol{v}_{-\cdot j}^{(\ell)}\right)\right)-u_{\cdot j}^{w}\left(v_{\cdot j}^{(\ell)} ; \boldsymbol{v}^{(\ell)}\right)\right] \text{\hfill .}
\end{array}\right.
\end{align*}

\section{Experiments}
\label{sec:exp}

In this section, we conduct empirical experiments on both synthetic data and real data, to validate the superiority of hybrid advertising model and the effectiveness of HRegNet. We run all the experiments on a server with both Central Processing Unit (CPU) and Graphics Processing Unit (GPU).

\textbf{Baseline methods.} HRegNet is compared with the following baselines:
\begin{itemize}
    \item \textbf{RegretNet} \cite{dutting2019optimal}. It generates the optimal mechanisms that satisfy near-DSIC and IR for traditional advertising models; 
    \item \textbf{JRegNet} \cite{zhang2024joint}. It is a neural network architecture for designing the optimal mechanisms of bundle-centric display; 
    \item \textbf{VCG} \cite{vickrey1961counterspeculation, clarke1971multipart, groves1973incentives}. It is a classic mechanism that satisfies DSIC and IR while maximizing social welfare.
    \item \textbf{Revised JRegNet}. Based on JRegNet, revised JRegNet aims to output a feasible hybrid allocation of independent stores and bundles. In detail, a dummy brand with value 0, is created to link to all stores, and generated ``bundles'' represent independent stores. A normalization-to-$C$ module is also added to limit the maximum number of winning bundles. Additionally, a weighted matrix (reflecting the CTR difference of independent stores and bundles) is used to converse the bundle allocation probability matrix to the advertiser allocation probability matrix in output layer. 
\end{itemize}

Note that in RegretNet, only stores can participate in the auctions. In JRegNet, stores and brands can bid in the auctions, but only the bundles can be winners and displayed in the slots. 

\textbf{Evaluation.} We use the following metrics to assess each mechanism: 
\begin{itemize}
    \item The empirical revenue: 
    $\frac{1}{L} \sum_{\ell=1}^{L} [\sum_{i=1}^{m}$ $ p_{i \cdot}^{w}(v^{(\ell)})+ \sum_{j=1}^{n} p_{\cdot j}^{w}(v^{(\ell)})].$
    \item The empirical social welfare:
    $\frac{1}{L} \sum_{\ell=1}^{L}(\sum_{i=1}^{m} \sum_{k=1}^{K} a_{i \cdot k}^{(\ell)} \theta_{k} v_{i \cdot}^{(\ell)}  + \sum_{j=1}^{n} \sum_{k=1}^{k} $ $a_{\cdot j k}^{(\ell)} \theta_{k} v_{\cdot j}^{(\ell)})$.
    \item The average empirical ex-post regret: 
    $\frac{1}{n+m} $ $(\sum_{i=1}^{m} \widehat{r g t}_{i \cdot} + \sum_{j=1}^{n} \widehat{r g t}_{\cdot j})$.
\end{itemize}

\subsection{Synthetic Data Experiments}

\textbf{Experimental details.}
For each experiment, the training set consists of $640000$ samples, while the test set contains $12800$ samples. During testing, we initialize $100$ different misreports for each advertiser in each auction sample and perform $200$ gradient ascent iterations on each misreport to update it. We then select the misreport that yields the maximum regret for each advertiser in each auction sample to calculate the advertiser's regret for that sample. Additionally, for each auction sample in every experiment, the joint matrix $\{\mathbf{1}_{ij}\}_{i\in M, j\in N}$ is randomly generated, and each store's quality factor $\alpha_i$ is drawn from $U[0.5, 1.5]$.



\textbf{Comparing HRegNet with baseline mechanisms in different settings.} We conduct a comprehensive evaluation of HRegNet against baseline methods across various settings, as detailed below:

\begin{enumerate}[leftmargin=*]
\item[(A)] $2$ stores, $2$ brands and $1$ slot with CTR $\boldsymbol{\theta}=(0.5)$. Each store's and brand's value is independently sampled from $U[0, 1]$.

\item[(B)] $3$ stores, $4$ brands and $3$ slots with CTRs $\boldsymbol{\theta}=(0.5, 0.3, 0.2)$. Each store's and brand's value is independently sampled from $U[0, 1]$.

\item[(C)] $3$ stores, $4$ brands and $4$ slots with CTRs $\boldsymbol{\theta}=(0.5, 0.3, 0.2, 0.1)$. Each store's and brand's value is independently sampled from $U[0, 1]$.

\item[(D)] $4$ stores, $4$ brands and $3$ slots with CTRs $\boldsymbol{\theta}=(0.5, 0.3, 0.2)$. Each store's and brand's value is independently sampled from $U[0, 1]$.
\end{enumerate}


\begin{table*}[ht]
\begin{tabular}{clllllll}
\hline
\multirow{2}{*}{\textbf{Method}} & \multirow{2}{*}{$\, \, C$} & {\multirow{2}{*}{{\begin{tabular}[c]{@{}c@{}}\multicolumn{1}{c}{\textbf{A: 2 $\times$ 2 $\times$ 1}}\\  rev \quad \quad sw \quad \quad rgt\end{tabular}}}} & {\multirow{2}{*}{{\begin{tabular}[c]{@{}c@{}}\multicolumn{1}{c}{\textbf{B: 3 $\times$ 4 $\times$ 3}}\\  rev \quad \quad sw \quad \quad rgt\end{tabular}}}} & {\multirow{2}{*}{{\begin{tabular}[c]{@{}c@{}}\multicolumn{1}{c}{\textbf{C: 3 $\times$ 4 $\times$ 4}} \\  rev 
\quad \quad sw \quad \quad rgt\end{tabular}}}} & {\multirow{2}{*}{{\begin{tabular}[c]{@{}c@{}}\multicolumn{1}{c}{\textbf{D: 4 $\times$ 4 $\times$ 3}} \\  rev 
\quad \quad sw \quad \quad rgt\end{tabular}}}} \\
                                 & \multicolumn{1}{c}{}                                                                                      & \multicolumn{1}{c}{}                                                                                      & \multicolumn{1}{c}{}                                                                                      & \multicolumn{1}{c}{}                                                                                      & \multicolumn{1}{c}{}                                                                                      & \multicolumn{1}{c}{}                                             \\ \hline \\ [-10pt] \hline

RegretNet                             &   \,  $-$ & 0.214$^{\phantom{\dagger}}$  0.311  \, \ \textless{}0.001                                                                                                                                                                                                      &   0.333$^{\phantom{\dagger}}$   0.496  \, \ \textless{}0.001                                                                                                      &     0.332$^{\phantom{\dagger}}$   0.497  \, \ \textless{}0.001                                                                                                     &     0.415$^{\phantom{\dagger}}$   0.608  \, \ \textless{}0.001                                                                                                                                            \\ \hline

JRegNet                             &   \,  $-$ & 0.245$^{\phantom{\dagger}}$  0.382  \, \ \textless{}0.001                                                                                                                                                                                                      &   0.607$^{\phantom{\dagger}}$   0.918  \, \ \textless{}0.001                                                                                                      &     0.621$^{\phantom{\dagger}}$   0.942  \, \ \textless{}0.001                                                                                                     &     0.684$^{\phantom{\dagger}}$   1.011  \, \ \textless{}0.001                                                                                                                                          \\ \hline

\multirow{3}{*}{VCG}                          

&   \! \, 1 & 0.196 \, \textbf{0.558}  \quad \,   $-$                                                                                                                                                                     &  0.496 \,  1.001  \quad \, \ \!  $-$                                                                                                      &     0.451  \, 1.026   \quad \, \ \! $-$                                                                                                     &    0.641   \,  1.078  \quad \, \ \! $-$ \\

&  \! \, 2 &  \, \ \! $-$ \quad \quad $-$  \quad \quad \! \!   $-$                                                                                                                                                                      &   0.560 \,  1.169  \quad \, \ \! $-$                                                                                                      &     0.545  \, 1.215  \quad \, \ \! $-$                                                                                                     &      0.722  \,  1.246   \quad \, \ \! $-$ 
\\

&   \! \, 3 & \, \ \! $-$ \quad \quad $-$  \quad \quad \! \!   $-$                                                                                                                                                                      &   0.537 \,  \textbf{1.223}   \quad \,  \!  $-$                                                                                                      &    0.537  \, \textbf{1.291}   \quad \,  \!  $-$                                                                                                     &      0.713   \,  \textbf{1.305}   \quad \,  \!  $-$

\\ \hline \multirow{3}{*}{Revised JRegNet}

& \! \, 1  & 0.315$^{\phantom{\dagger}}$  0.439  \,  \textless{}0.001                                                                                                                                                                                                      &   0.606$^{\phantom{\dagger}}$   0.817  \, \ \textless{}0.001                                                                                                      &     0.609$^{\phantom{\dagger}}$   0.818  \, \ \textless{}0.001                                                                                                     &     0.670$^{\phantom{\dagger}}$   0.873  \, \ \textless{}0.001

\\ 

                       & \! \, 2  & \, \ \! $-$ \quad \quad $-$  \quad \quad \! \!   $-$                                                                                                                                                                                                          &   0.717$^{\phantom{\dagger}}$   0.988  \, \ \textless{}0.001                                                                                                      &     0.724$^{\phantom{\dagger}}$   1.002  \, \ \textless{}0.001                                                                                                     &     0.760$^{\phantom{\dagger}}$   1.026  \, \ \textless{}0.001   \\ 

                       &  \! \, 3 & \, \ \! $-$ \quad \quad $-$  \quad \quad \! \!   $-$                                                                                                                                                                          & 0.718$^{\phantom{\dagger}}$  1.016 \, \ \textless{}0.001                                                                                 & 0.751$^{\phantom{\dagger}}$     1.057  \, \  \textless{}0.001                                                                                 & 0.788$^{\phantom{\dagger}}$     1.084  \, \  \textless{}0.001 
                       \\ 

                               \hline

        \multirow{3}{*}{HRegNet}

& \! \, 1  & \textbf{0.347$^{\dagger}$}  0.476  \,  \textless{}0.001                                                                                                                                                                                                      &   0.663$^{\dagger}$   0.883  \, \ \textless{}0.001                                                                                                      &     0.670$^{\dagger}$   0.892  \, \ \textless{}0.001                                                                                                     &     0.755$^{\dagger}$   0.966  \, \ \textless{}0.001

\\ 

                       & \! \, 2  & \, \ \! $-$ \quad \quad $-$  \quad \quad \! \!   $-$                                                                                                                                                                                                          &   0.750$^{\dagger}$   1.029  \, \ \textless{}0.001                                                                                                      &     0.765$^{\dagger}$   1.060  \, \ \textless{}0.001                                                                                                     &     0.853$^{\dagger}$   1.123  \, \ \textless{}0.001   \\ 

                       &  \! \, 3 & \, \ \! $-$ \quad \quad $-$  \quad \quad \! \!   $-$                                                                                                                                                                         & \textbf{0.770$^{\dagger}$}  1.067 \, \textless{}0.001                                                                                 & \textbf{0.791$^{\dagger}$}     1.111  \, \  \textless{}0.001                                                                                 & \textbf{0.876$^{\dagger}$}     1.164  \,   \textless{}0.001 
                       \\ 

                               \hline

\end{tabular}
\caption{The experimental results of Settings A to D. The parameter $C$ denotes the maximum number of winning bundles. The symbol``${\dagger}$'' indicates that the revenue shows a statistically significant improvement compared to other methods within the same $C$, as determined by a paired $t$-test at $p < 0.05$. }
\label{tbl:1}
\end{table*}

\begin{table*}[ht]
\begin{tabular}{cllllll}
\hline
\multirow{2}{*}{\textbf{Method}} & \multirow{2}{*}{$\, \, C$} & {\multirow{2}{*}{{\begin{tabular}[c]{@{}c@{}}\multicolumn{1}{c}{\textbf{Uniform}}\\  rev \quad \quad sw \quad \quad rgt\end{tabular}}}} & {\multirow{2}{*}{{\begin{tabular}[c]{@{}c@{}}\multicolumn{1}{c}{\textbf{Normal}}\\  rev \quad \quad sw \quad \quad rgt\end{tabular}}}} & {\multirow{2}{*}{{\begin{tabular}[c]{@{}c@{}}\multicolumn{1}{c}{\textbf{Lognormal}} \\  rev 
\quad \quad sw \quad \quad rgt\end{tabular}}}} \\
                                 & \multicolumn{1}{c}{}                                                                                      & \multicolumn{1}{c}{}                                                                                      & \multicolumn{1}{c}{}                                                                                      & \multicolumn{1}{c}{}                                                                                      & \multicolumn{1}{c}{}                                                                                      & \multicolumn{1}{c}{}                                             \\ \hline \\ [-10pt] \hline

RegretNet                             &   \,  $-$ & 0.333$^{\phantom{\dagger}}$   0.496  \, \ \textless{}0.001                                                                                                                                                                                                      &   0.334$^{\phantom{\dagger}}$   0.498  \, \ \textless{}0.001                                                                                                      &     0.296$^{\phantom{\dagger}}$   0.453  \, \ \textless{}0.001                                                                                                                                              \\ \hline

JRegNet                             &   \,  $-$ & 0.607$^{\phantom{\dagger}}$   0.918  \, \ \textless{}0.001                                                                                                                                                                                                      &   0.630$^{\phantom{\dagger}}$   0.937  \, \ \textless{}0.001                                                                                                      &     0.559$^{\phantom{\dagger}}$   0.882 \, \ \textless{}0.001                                                                                                                                            \\ \hline

\multirow{3}{*}{VCG}                          

&   \! \, 1 & 0.496 \,  1.001 \quad \, \, $-$                                                                                                                                                                     &   0.502 \,  0.974  \quad \, \, $-$                                                                                                      &    0.452  \, 0.925  \quad \, \, $-$                                                                                                     \\

&  \! \, 2 & 0.560 \,  1.169  \quad \, \, $-$                                                                                                                                                                     &   0.570 \,  1.135  \quad \, \, $-$                                                                                                      &     0.510  \, 1.081   \quad \, \, $-$                                                                                                     
\\

&   \! \, 3 & 0.537 \,  \textbf{1.223}  \quad \, \ $-$                                                                                                                                                                     &   0.553 \,  \textbf{1.189}  \quad \, \ $-$                                                                                                      &     0.490  \, \textbf{1.132}   \quad \, \ $-$

\\ \hline

\multirow{3}{*}{Revised JRegNet}

                &   \! \, 1   & 0.606$^{\phantom{\dagger}}$   0.817  \, \ \textless{}0.001                                                                                                                                                                                                      &   0.605$^{\phantom{\dagger}}$   0.822  \, \ \textless{}0.001                                                                                                      &     0.539$^{\phantom{\dagger}}$   0.762  \, \ \textless{}0.001                                                                                                    
\\ 

                       & \! \, 2  & 0.717$^{\phantom{\dagger}}$   0.988  \, \ \textless{}0.001                                                                                                                                                                                                      &   0.715$^{\phantom{\dagger}}$   0.992 \, \ \textless{}0.001                                                                                                      &     0.634$^{\phantom{\dagger}}$   0.927  \, \ \textless{}0.001

\\ 

                       &  \! \, 3 & 0.718$^{\phantom{\dagger}}$  1.016 \, \ \textless{}0.001                                                                                                                                                                & 0.743$^{\phantom{\dagger}}$  1.042 \, \ \textless{}0.001                                                                                 & 0.645$^{\phantom{\dagger}}$     0.955  \, \  \textless{}0.001      
                                                            \\ \hline

        \multirow{3}{*}{HRegNet}

                &   \! \, 1   & 0.663$^{\dagger}$   0.883  \, \ \textless{}0.001                                                                                                                                                                                                      &   0.659$^{\dagger}$   0.884  \, \ \textless{}0.001                                                                                                      &     0.593$^{\dagger}$   0.829  \, \ \textless{}0.001                                                                                                    
\\ 

                       & \! \, 2  & 0.750$^{\dagger}$   1.029 \, \ \textless{}0.001                                                                                                                                                                                                      &   0.753$^{\dagger}$   1.032 \, \ \textless{}0.001                                                                                                      &     0.675$^{\dagger}$   0.963  \, \ \textless{}0.001

\\ 

                       &  \! \, 3 & \textbf{0.770$^{\dagger}$}  1.067 \,  \textless{}0.001                                                                                                                                                                & \textbf{0.773$^{\dagger}$}  1.067 \, \textless{}0.001                                                                                 & \textbf{0.690$^{\dagger}$}     1.002  \,   \textless{}0.001      
                                                            \\ \hline

\end{tabular}
\caption{The experimental results of different value distributions. The experimental setting is Setting B. The parameter $C$ denotes the maximum number of winning bundles. The symbol``${\dagger}$'' indicates that the revenue shows a statistically significant improvement compared to other methods within the same $C$, as determined by a paired $t$-test at $p < 0.05$. }
\label{tbl:2}
\end{table*}

\begin{table*}[ht]
\begin{tabular}{cllllll}
\hline
\multirow{2}{*}{\textbf{Method}} & \multirow{2}{*}{$\, \, C$} & {\multirow{2}{*}{{\begin{tabular}[c]{@{}c@{}}\multicolumn{1}{c}{\textbf{E1}}\\  rev \quad \quad \quad sw \quad \quad ru\end{tabular}}}} & {\multirow{2}{*}{{\begin{tabular}[c]{@{}c@{}}\multicolumn{1}{c}{\textbf{E2}}\\  rev \quad \quad sw \quad \quad ru\end{tabular}}}} & {\multirow{2}{*}{{\begin{tabular}[c]{@{}c@{}}\multicolumn{1}{c}{\textbf{E3}} \\  rev 
\quad \quad sw \quad \quad ru\end{tabular}}}} \\
                                 & \multicolumn{1}{c}{}                                                                                      & \multicolumn{1}{c}{}                                                                                      & \multicolumn{1}{c}{}                                                                                      & \multicolumn{1}{c}{}                                                                                      & \multicolumn{1}{c}{}                                                                                      & \multicolumn{1}{c}{}                                             \\ \hline \\ [-10pt] \hline
                            RegretNet                             &   \,  $-$ & 10.815$^{\phantom{\dagger}}$ \, \! \! 18.072  \,  \textless{}0.065                                                                                                                                                                      &   9.678$^{\phantom{\dagger}}$ \ \! \!  16.023  \, \textless{}0.065                                                                                                     &     11.441$^{\phantom{\dagger}}$   19.310   \, \textless{}0.065   
 \\ \hline 
JRegNet                             &   \,  $-$ & 19.067$^{\phantom{\dagger}}$ \, \! \! 32.711  \,  \textless{}0.065                                                                                                                                                                     &   18.124$^{\phantom{\dagger}}$   31.806  \, \textless{}0.065                                                                                                     &     19.628$^{\phantom{\dagger}}$   35.006   \, \textless{}0.065   
 \\ \hline 
\multirow{5}{*}{VCG}                          

&  \! \, 1 & 10.913$^{\phantom{\dagger}}$ \, \! \! 37.249  \quad \ \ $-$                                                                                                                                                                     &   10.645$^{\phantom{\dagger}}$   35.547 \quad \ \ $-$                                                                                                      &    10.835$^{\phantom{\dagger}}$  37.828  \quad \ \ $-$                                                                                                     \\

& \! \, 2 & 17.676$^{\phantom{\dagger}}$ \, \! \! 45.274  \quad \ \ $-$                                                                                                                                                                     &   16.842$^{\phantom{\dagger}}$   43.292 \quad \ \ $-$                                                                                                      &      17.499$^{\phantom{\dagger}}$  45.839  \quad \ \ $-$                                                                                                     
\\

&  \! \, 3 & 22.109$^{\phantom{\dagger}}$ \, \! \! 50.782  \quad \ \ $-$                                                                                                                                                                     &   20.913$^{\phantom{\dagger}}$  48.752 \quad \ \ $-$                                                                                                      &    21.603$^{\phantom{\dagger}}$  51.395 \ \quad \ $-$

\\

&  \! \, 4 & 23.580$^{\phantom{\dagger}}$ \, \! \! 53.794  \quad \ \ $-$                                                                                                                                                                     &   22.567$^{\phantom{\dagger}}$  51.821  \quad \ \ $-$                                                                                                      &    23.143$^{\phantom{\dagger}}$  54.373   \quad \ \ $-$ 

\\

& \! \,  5 & 23.566$^{\phantom{\dagger}}$ \, \! \! \textbf{54.823}  \, \!  \! \! \! \ \ $-$                                                                                                                                                                     &  22.639$^{\phantom{\dagger}}$   \textbf{52.956}  \, \!  \! \! \! \ \ $-$                                                                                                      &    23.201$^{\phantom{\dagger}}$  \textbf{55.356}   \, \!  \! \! \! \ \ $-$

\\ \hline

\multirow{5}{*}{Revised JRegNet}                    

                       & \! \, 1 & 11.676$^{\phantom{\dagger}}$ \, \! \! 16.937 \, \    \textless{}0.065                                                                                                                                                              & 9.999$^{\phantom{\dagger}}$ \ \! \! 15.168 \, \ \textless{}0.065                                                                                & 11.994$^{\phantom{\dagger}}$     17.765 \, \  \textless{}0.065     
\\ 

                       & \! \, 2  & 15.261$^{\phantom{\dagger}}$ \, \! \! 23.221  \, \ \textless{}0.065                                                                                                                                                                                                    &   14.612$^{\phantom{\dagger}}$   22.260  \, \ \textless{}0.065                                                                                                      &     15.697$^{\phantom{\dagger}}$   23.504  \, \ \textless{}0.065    
                       \\ 

                       & \! \, 3  & 18.213$^{\phantom{\dagger}}$ \, \! \! 27.953  \, \ \textless{}0.065                                                                                                                                                                                                     &   18.527$^{\phantom{\dagger}}$   28.920 \, \ \textless{}0.065                                                                                                       &     18.918$^{\phantom{\dagger}}$   28.553 \, \ \textless{}0.065  \\ 

                       & \! \, 4  & 22.495$^{\phantom{\dagger}}$ \, \! \! 36.118  \, \ \textless{}0.065                                                                                                                                                                                                   &   22.919$^{\phantom{\dagger}}$  36.375  \, \ \textless{}0.065                                                                                                      &     23.902$^{\phantom{\dagger}}$   37.936  \, \  \textless{}0.065  \\ 

                       & \! \, 5  & 25.010$^{\phantom{\dagger}}$ \, \! \! 40.276  \, \ \textless{}0.065                                                                                                                                                                                                     &   23.011$^{\phantom{\dagger}}$   37.704  \, \  \textless{}0.065                                                                                                    &     23.721$^{\phantom{\dagger}}$   37.633  \, \    \textless{}0.065
                       \\ \hline
        \multirow{5}{*}{HRegNet}                    

                       & \! \, 1 & 20.092$^{\dagger}$ \, \! \! 29.990 \, \ \! \textless{}0.065                                                                                                                                                             &                                19.099$^{\dagger}$     29.126 \, \  \textless{}0.065                                                    & 

                       20.936$^{\dagger}$ 31.536 \, \ \textless{}0.065 
\\ 

                       & \! \, 2  & 22.253$^{\dagger}$ \, \! \! 35.222  \, \ \! \textless{}0.065                                                                                                                                                                                                    &  22.073$^{\dagger}$   34.132  \, \ \textless{}0.065                                                                                                         &       23.560$^{\dagger}$   36.042  \, \ \textless{}0.065 
                       \\ 

                       & \! \, 3  & 23.794$^{\dagger}$ \, \! \! 38.907  \, \ \! \textless{}0.065                                                                                                                                                                                                    &     22.312$^{\dagger}$   35.681 \, \ \textless{}0.065                                                                                                &     24.774$^{\dagger}$   40.386  \, \ \textless{}0.065     \\ 

                       & \! \, 4  & 24.074$^{\dagger}$ \, \! \! 38.024  \, \ \! \textless{}0.065                                                                                                                                                                                                     &   23.948$^{\dagger}$   38.591  \, \  \textless{}0.065                                                                                                   &    26.242$^{\dagger}$   42.351  \, \ \textless{}0.065      \\ 

                       & \! \, 5  & \textbf{25.705}$^{\dagger}$ \,   40.666  \, \ \textless{}0.065                                                                                                                                                                                                     &   \textbf{24.185}$^{\dagger}$   39.850 \,  \textless{}0.065                                                                                                     &    \textbf{26.635}$^{\dagger}$   42.511 \,  \textless{}0.065 
                       \\ \hline

\end{tabular}
\caption{The experimental results for real online auction data on the test set. 
The parameter $C$ denotes the maximum total number of winning bundles. The symbol``${\dagger}$'' indicates that the revenue shows a statistically significant improvement compared to other methods within the same $C$, as determined by a paired $t$-test at $p < 0.05$.}
\label{tbl:999}
\end{table*}

The experimental results for settings A to D are presented in Table \ref{tbl:1}. 
First, considering the experimental result of RegretNet, JRegNet, and HRegNet, 
we find that, even though we limit the number of winning bundles $C$ to be 1 and the regret levels to be below 0.001, HRegNet consistently achieves the highest revenue. This indicates that the hybrid advertising model generates more revenue than both the joint advertising model and the traditional advertising model. 
Additionally, Table \ref{tbl:1} also shows that in the hybrid advertising model, HRegNet significantly outperforms VCG and revised JRegNet in revenue across all settings and values of $C$, despite a certain level of social welfare loss compared with VCG. 
This demonstrates the effectiveness of the HRegNet-generated mechanism in hybrid advertising.


\textbf{Different value distributions}. 
To further evaluate the performance of the hybrid advertising model and the mechanisms generated by HRegNet, we conduct experiments under various value distributions, using not only a uniform distribution $U[0, 1]$ but also a truncated normal distribution $N(0.5,0.16)$ with $v \in [0,1]$ and a truncated log-normal distribution $LN(0.1,1.69)$ with $v \in [0,1]$ in setting B. The experimental results are shown in Table \ref{tbl:2}.



From Table \ref{tbl:2}, we observe that HRegNet achieves significantly higher revenue than RegretNet and JRegNet across three different value distributions while maintaining a regret level below $0.001$. 
Additionally, HRegNet also consistently outperforms VCG and revised JRegNet for all three distributions, demonstrating its robustness in handling various value distributions.


\subsection{Real-World Data Experiments}
In this subsection, we use real log data from an e-commerce platform's online ad auctions to train and test HRegNet. In the online ad auction scenario, when a user's search request arrives, the process involves ad recall, rough ranking, and fine ranking. Up to 10 bundles, including a maximum of 10 stores and 10 brands, bid for 5 ad slots during the auction stage. We evaluate our model in this setting, where 10 stores and 10 brands compete for 5 ad slots. The log data we utilize includes the identifier and bid for each advertiser, the predicted CTR for each bundle and each independent store, and the joint relationship between stores and brands. To accommodate the varying number of advertisers in the initial dataset, we adjust some auction samples by trimming or padding to ensure that each sample includes 10 stores and 10 brands, padding with advertisers assigned a value per click of zero.

Additionally, because the utility of advertisers in real data differs significantly, using regret alone cannot accurately measure the degree of DSIC. Therefore, in our real data experiments, we evaluate the degree of DSIC using the ratio of achievable utility increase (regret) through misreporting to the utility obtained from truthful reporting:
$r u \coloneq \frac{1}{L} \sum_{\ell=1}^{L}[(\sum_{i=1}^{m} r g t_{i \cdot}^{(\ell)}+\sum_{j=1}^{n} r g t_{\cdot j}^{(\ell)}) /(\sum_{i=1}^{m} \mu_{i \cdot}^{(\ell)}+ \sum_{j=1}^{n} \mu_{\cdot j}^{(\ell)})],$
where $r g t_{i \cdot} \coloneq \max_{v_{i \cdot}^{\prime} \in V_{i \cdot}} u_{i \cdot} (v_{i \cdot}; (v_{i \cdot}^{\prime}, \boldsymbol{v}_{-i \cdot})) - u_{i \cdot} (v_{i \cdot}; \boldsymbol{v})$  (a similar definition applies to brand $j$).

We conduct three groups of real-data experiments: E1 uses auction data from December 24–25, 2024 for training and December 26, 2024 for testing; E2 uses data from December 25–26 for training and December 27 for testing; and E3 uses data from December 26–27 for training and December 28 for testing. Each real data experiment utilizes 6,912 real ad auction samples for testing.




As shown in Table \ref{tbl:999}, 
for all the real data experiments, HRegNet significantly outperforms RegretNet and JRegNet in terms of revenue. Furthermore, Table \ref{tbl:999} demonstrates that HRegNet generates significantly higher revenue than VCG and revised JRegNet across all real data experiments and values of $C$. These results confirm the effectiveness of HRegNet in real-world hybrid auction settings.

Furthermore, HRegNet can be deployed in real ad scenarios. Using a server with 48 CPUs and 1 GPUs, completing 30000 offline training iterations for HRegNet takes approximately 2 to 4 hours. In contrast, invoking the trained HRegNet for forward propagation to derive allocation and payment matrices is extremely quick, taking about 1 to 3 milliseconds per auction sample, ensuring that online decision-making remains unaffected.


\section{Conclusion}
\label{sec:conc}


In this paper, we propose a novel hybrid advertising model where an ad slot is ultimately allocated to either an independent store or a bundle comprising a store and a brand. 
Then we introduce HRegNet, a neural network architecture, for generating the optimal hybrid auction mechanisms that satisfy near-DSIC and IR. 
Subsequently, we perform numerical experiments on both synthetic data and real data, demonstrating that the mechanism generated by HRegNet significantly outperforms other baseline methods.

Note that in online retrieval-based advertising systems,
the platform sequentially performs ad recall, rough ranking, and fine ranking when a user searches. Our work mainly focuses on the fine ranking stage; however, to deploy the hybrid ad model online, some adjustments are also needed in the recall stage. For example, during the recall stage, both independent store ads and bundle ads need to be recalled. 
In the future, investigating dynamic auction mechanisms for hybrid advertising in multi-query settings is promising.

\section*{Acknowledgments}

Qi Qi is the corresponding author. This work was supported by National Natural Science Foundation of China NO. 62472428, the Fundamental Research Funds for the Central Universities, and the Research Funds of Renmin University of China (No. 22XNKJ07, No. 23XNH028), Major Innovation \& Planning Interdisciplinary Platform for the “Double-First Class” Initiative, Renmin University of China, and Meituan Inc. Fund.






\newpage
\bibliographystyle{plain}
\bibliography{references}

\end{document}